\documentclass[runningheads]{llncs}

\usepackage{graphicx}
\usepackage{xcolor}
\usepackage{multirow}
\usepackage{float}
\begin{document}
\title{Predicting isocitrate dehydrogenase mutation status in glioma using structural brain networks and graph neural networks }
\titlerunning{Predicting IDH status in glioma using brain networks and GNN}
%
\author{Yiran Wei$^\star$\inst{1} \and Yonghao Li\thanks{Equal contribution}\inst{2} \and Xi Chen\inst{3}\and Carola-Bibiane Schönlieb\inst{4} \and  Chao Li\thanks{Corresponding author}\inst{1,4} \and  Stephen J. Price\inst{1}} 
%
\authorrunning{Y. Wei et al.}
%
\institute{Department of Clinical Neuroscience, University of Cambridge  
\and School of Biological Sciences,  Nanyang Technological University
\and Department of Computer Science, University of Bath
\and Department of Applied Mathematics and Theoretical Physics, University of Cambridge}
\maketitle              
\begin{abstract}
Glioma is a common malignant brain tumor with distinct survival among patients. The isocitrate dehydrogenase (IDH) gene mutation provides critical diagnostic and prognostic value for glioma. It is of crucial significance to non-invasively predict IDH mutation based on pre-treatment MRI. Machine learning/deep learning models show reasonable performance in predicting IDH mutation using MRI. However, most models neglect the systematic brain alterations caused by tumor invasion, where widespread infiltration along white matter tracts is a hallmark of glioma. Structural brain network provides an effective tool to characterize brain organisation, which could be captured by the graph neural networks (GNN) to more accurately predict IDH mutation.

Here we propose a method to predict IDH mutation using GNN, based on the structural brain network of patients. Specifically, we firstly construct a network template of healthy subjects, consisting of atlases of edges (white matter tracts) and nodes (cortical/subcortical brain regions) to provide regions of interest (ROIs). Next, we employ autoencoders to extract the latent multi-modal MRI features from the ROIs of edges and nodes in patients, to train a GNN architecture for predicting IDH mutation.  The results show that the proposed method outperforms the baseline models using the 3D-CNN and 3D-DenseNet. In addition, model interpretation suggests its ability to identify the tracts infiltrated by tumor, corresponding to clinical prior knowledge. In conclusion, integrating brain networks with GNN offers a new avenue to study brain lesions using computational neuroscience and computer vision approaches.

\end{abstract}
\section{Introduction}

\subsection{Significance of predicting IDH mutational status}
Gliomas are common malignant brain tumors with various prognosis~\cite{ostrom2016cbtrus}. The mutation status of isocitrate dehydrogenase (IDH) genes is one of the most important biomarkers for the diagnosis and prognosis of gliomas, where IDH mutants tend to have a better prognosis than IDH wild-types~\cite{yan2009idh1}. Due to the crucial value in clinical practice, IDH mutations have been established as one of the landmark molecular markers for glioma patients, recommended by the World Health Organization classification of tumors of the Central Nervous System for routine 
assessment in glioma patients~\cite{louis20162016}. 

Currently, the most widely used approaches to determine IDH mutation status, i.e., immunohistochemistry and targeted gene sequencing, rely on tumor samples~\cite{louis20162016}, which therefore cannot be assessed on those patients who are not suitable for tumor resection or biopsy. Further, as the assays usually are time-consuming and expensive, they are not available in some institutions. 

Meanwhile, the radiogenomic approach has shown promise in predicting molecular markers based on radiological images. Mounting evidence has supported the feasibility of predicting IDH mutation status using the pre-operative MRI~\cite{choi2021fully,hyare2019modelling,liang2018multimodal}. The most commonly used MRI sequences include pre-contrast T1, post-contrast T1, T2, and T2-weighted-Fluid-Attenuated Inversion Recovery (FLAIR). Integrating the quantitative information from multi-modal MRI  promises to provide a non-invasive approach to characterize glioma and predict IDH mutations for better treatment planning and prognostication~\cite{li2019characterizing,li2019multi}.

\subsection{Brain structural networks}

The tissue structure of the human brain is divided into grey matter and white matter. The grey matter, located on the brain surface, constitutes the cerebral cortex and can be parcelled into cortical/subcortical regions based on cortical gyri and sulci. The parcellation offers a more precise association between brain function with cortical structure. The white matter of the cerebral cortex contains the connecting axons among the cortical/subcortical regions.  The structural network of the brain is a mathematical simplification of the connectivity of the cortical/subcortical regions~\cite{bullmore2011brain}, where the nodes represent the cortical/subcortical regions and the edges are defined as connecting white matter tracts.

Accumulating research of structural brain networks 
has reported significance in neuropsychiatric diseases, including stroke, traumatic brain injury, and brain tumors~\cite{fagerholm2015disconnection,salvalaggio2020post,liu2020altered,wei2021structural}. On the other hand, evidence shows that glioma cells tend to invade along the white matter pathway~\cite{wang2019invasion} and infiltrate the whole brain~\cite{stoecklein2020resting,wei2021structural}. Therefore, investigating structural brain networks could offer a tool to investigate glioma invasion on both tumor core and normal-appearing brain regions. Further, a previous study revealed that IDH mutations could be associated with different invasive phenotypes of glioma~\cite{rice2017lessp}. To this end, we hypothesize that employing the structural brain networks could provide value for predicting IDH mutation status. In particular, with prior  knowledge of brain structure and anatomy incorporated, a more robust prediction model could be achieved.

\subsection{Graph neural networks}
The graph neural networks (GNN) is a branch of deep learning, specialized in data formats of irregular structures, such as varying numbers of edges and random orders of nodes in graph data~\cite{morris2019weisfeiler}. Unlike the traditional convolutional neural networks (CNN) that convolute elements one by one in the grid data, the GNN aggregate information into nodes from their neighbors and simultaneously learns a representation of the whole graph. By employing the GNN on structural brain networks, the topological information contained in the structural brain networks could be effectively explored, which would consequently incorporate the prior knowledge of brain organization and perceive the critical information of tumor invasion at the whole-brain level.

\subsection{Related work}
Current methods of predicting IDH mutation status include radiomics/machine learning-based, deep learning-based, or a combination of both. 
Radiomics/machine learning-based methods extract high dimensional handcrafted features from the MRIs, e.g., tumor intensity, shape, texture, etc., to train machine learning prediction models of molecular markers, tumor grades, or patient survival~\cite{hyare2019modelling}. 
Deep learning-based approaches provide end-to-end model without pre-defined imaging features in the prediction tasks~\cite{liang2018multimodal}. Some other methods integrated the radiomic features into a deep neural network to enhance prediction performance~\cite{choi2021fully}. Albeit reasonable prediction accuracy, most of these methods are mainly driven by the computer vision tasks, without considering the systematic alteration of the brain organization during tumor invasion. Incorporating the prior knowledge from the neuroscience field shows promise to improve the prediction model.

\subsection{Proposed methods}
Here we propose an approach of using GNN to predict IDH mutation status, based on the structural brain networks generated from multi-model MRI and prior human brain atlases. Our contributions include: 

\begin{itemize}
   
    \item A method to incorporate the prior knowledge of brain atlases with the anatomical MRI to generate structural brain networks.
     
    \item A novel architecture of GNN with specialized graph convolutional operator for aggregating multi-dimensional latent features of the multi-model MRI.
    
    \item To our best knowledge, this is the first study that leverages GNN on the multi-modal MRI to predict the IDH mutation status of glioma.

\end{itemize}

\section{Methods}

\subsection{Datasets}
This study included the pre-operative multi-modal MRI (pre-contrast T1, post-contrast T1, T2, and FLAIR) of 389 glioma patients. MRI images of 274 patients were downloaded from The Cancer Imaging Archive (TCIA) website \cite{tcgalgg,tcgagbm,ivygap}, whereas 115 patients were available from an in-house cohort. 17 of 389 patients who have missing IDH mutation status or incomplete MRI modalities were excluded. For the included patients, 103  patients are IDH mutant, and 269 are IDH wild-type.

\subsection{Imaging pre-processing}
We processed the multi-modal MRI following a standard pipeline~\cite{bakas2017advancing}. Firstly, the T1, T2, and FLAIR were co-registered to the post-contrast T1 using the FMRIB's Linear Image Registration Tool \cite{jenkinson2001global}. Then, brain extraction was performed on all MRI modalities to remove the skull using Brain Extraction Tool in the FMRIB Software Library (FSL) \cite{jenkinson2012fsl,smith2002fast}. We also performed histogram matching~\cite{nyul2000new} and voxel smoothing with SUSAN noise reduction~\cite{smith1997susan}. A neurosurgeon and a researcher performed manual correction of brain masks, cross-validated using DICE score. Finally, all modalities were non-linearly co-registered using the Advanced Normalization Tools (ANTs)~\cite{avants2009advanced} to the MNI152 standard space, i.e., MNI-152-T1-2MM-brain provided by the FSL (Figure~\ref{Fig:brainnetworks}A).

\subsection{Constructing patient structural brain networks}

\subsubsection{Brain network template}
\begin{figure}[H]
\begin{center}
\centerline{\includegraphics[width=\columnwidth]{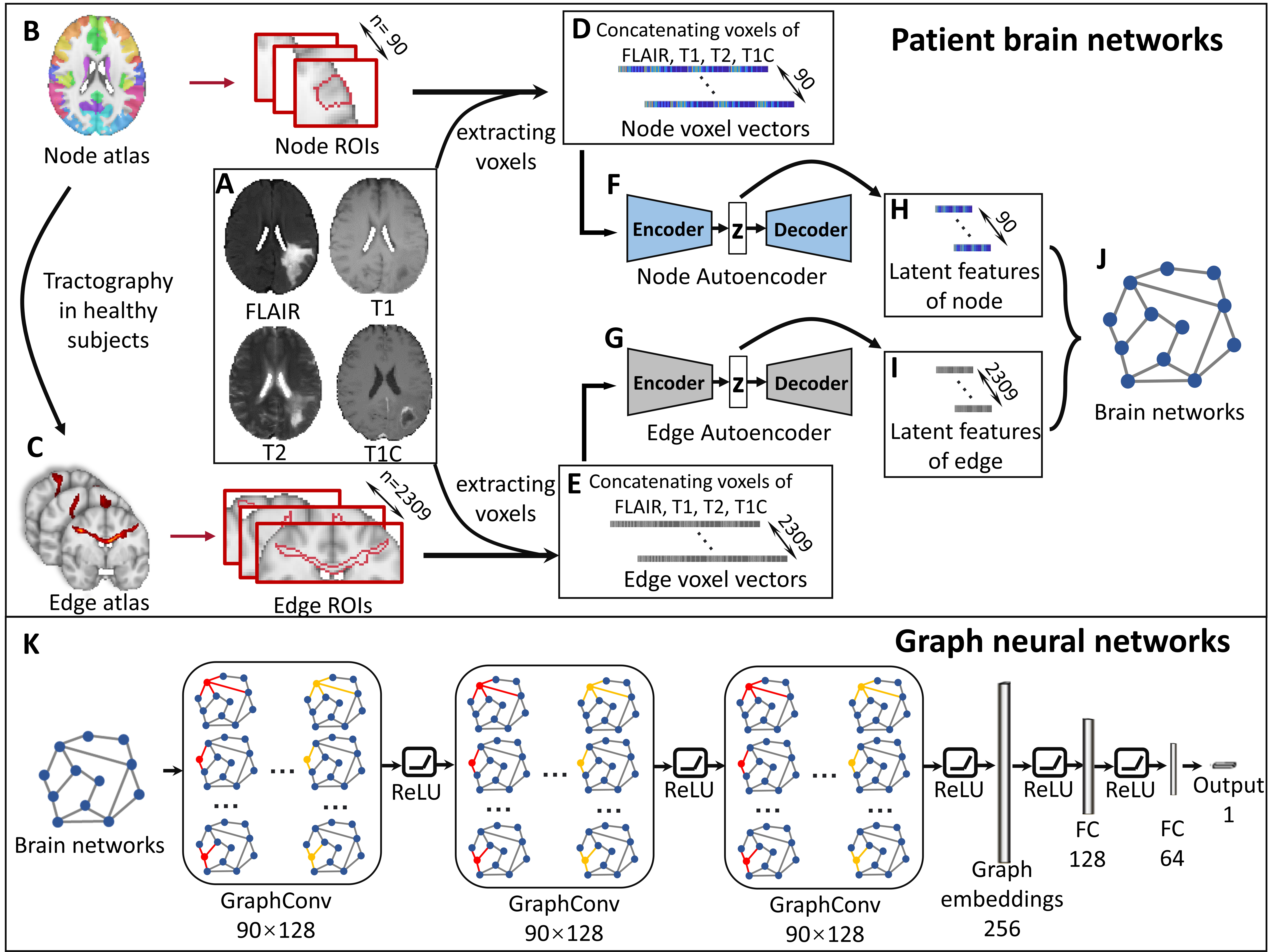}}
\caption{Study workflow. Upper: the pipeline of constructing patient brain networks. \textbf{A}: Patient MRIs are pre-processed and co-registered to the atlas space.  \textbf{B}: The AAL atlas of 90 ROIs is used as the node atlas. \textbf{C}: The edge atlas is generated from performing tractography among the 90 ROIs on the diffusion MRI of healthy subjects. \textbf{D \& E} Multi-modal MRI voxels within the node/edge ROIs are extracted and concatenated to voxel vectors to characterize the node/edge.  90 node were from AAL atlas while 2309 edges are the edges that exist in 9 of 10 healthy subjects in tractography. \textbf{F \& G}: Two autoencoders are trained using edge and node voxel vectors. \textbf{H \& I}: Encoders of trained autoencoders are used to extract the low dimensional latent features $z$ from the high dimensional node/edge voxels vector, respectively. \textbf{J} Latent node/edge features are then rearranged into graph format as the input of the GNN. \textbf{K} Graph convolutional neural networks consist of three hidden graph convolutional layers, one graph embedding layer, and two fully-connected (FC) layers.}
\label{Fig:brainnetworks}
\end{center}
\end{figure}

We leveraged the brain network template derived from healthy subjects to construct brain networks in lesioned brains~\cite{salvalaggio2020post}. First, we used the prior brain atlases in healthy subjects as the template of brain networks, generating regions of interest (ROIs) for characterizing the brain networks in patients based on multi-modal MRI. Specifically, we used the Automated Anatomical Labelling (AAL) atlas~\cite{tzourio2002automated} as the node ROIs (Figure~\ref{Fig:brainnetworks}B), which includes 90 brain cortical and subcortical regions. Further, we generated a brain connectivity atlas from ten healthy subjects scanned by high-resolution diffusion MRI to derive the edge ROIs of the structural brain networks (Figure~\ref{Fig:brainnetworks}C). We used a similar approach of generating brain connectivity atlas with~\cite{fagerholm2015disconnection,wei2021quantifying}. In brief, firstly, pairwise tractography among the 90 regions of AAL atlas was performed in healthy subjects, then the resultant tract pathways were co-registered to the MNI152 standard space. Next, the corresponding tracts of all healthy subjects were averaged for each edge between two nodes. Finally, the top 5\% voxels of the tract density were retained and binarized to generate robust edge ROIs. The generated edge atlas is shown in Figure~\ref{Fig:brainnetworks}C.

\subsubsection{Latent features of nodes and edges from autoencoders}
MRI voxels within the ROIs of the node or edge atlases across the whole brain were extracted and then concatenated to voxel vectors (Figure~\ref{Fig:brainnetworks}D \& E). We then used two autoencoders to extract the latent features from the voxel vectors of node and edge, respectively. Vector size was set as 2500 (voxels) $\times$ 4 (modalities) = 10000. For edges and nodes with few voxels, the vectors were padding with zeros. The patient cohort was shuffled and split into a 80:20 ratio for training and testing data. Two autoencoders were trained by edge and node voxel vectors of the training data (Figure~\ref{Fig:brainnetworks}F \& G). Finally, the latent features of edge or node voxels were derived, with the dimension of the edge or node vectors substantially decreased from 10000 to 12 (Figure~\ref{Fig:brainnetworks}H \& I). The 12 latent features  were used as the input of the GNN (Figure~\ref{Fig:brainnetworks}J). Logistic sigmoid function was applied as transfer function for both encoder and decoder. L2 regularization with coefficient of 0.001 was used. 'msesparse' was set as the loss function.

\subsection{Predicting IDH mutation status using GNN}
The patient brain networks constructed above were used to train the GNN, with the multi-modal MRI latent features as inputs. In addition to the 80:20 ratio of training and testing data, training data was split again into an 80:20 ratio for cross-validation. The proposed GNN consist of three graph convolutional layers similar to the one defined in~\cite{morris2019weisfeiler},  one node to graph embedding layers, and two fully connected feed forward layers (Figure ~\ref{Fig:brainnetworks}K). We used a binary cross-entropy loss, while the optimization was done using Adam optimizer.  

The graph convolutional operator is defined as follow:
\begin{equation}
\mathbf{x_i^{\prime}} =  \mathbf{\Theta}_1 \mathbf{x}_i +  \sum_{z=1}^{Z} \mathbf{\Theta}_2 \sum_{j\in\mathcal{N}(i)} e_{j,i,z} \cdot \mathbf{x}_{j}
\label{Eq:graphconv}
\end{equation}

Where $\mathbf{x_i^{\prime}}$ denotes the features of node $i$ after convolution, $ \mathbf{\Theta}_1$ and $ \mathbf{\Theta}_2$ denote the trainable network weights. $\cdot$ is the multiply operator.  $e_{j,i,z}$ represents the $z$th edge feature from source node $j$ to target node $i$. $j\in\mathcal{N}(i)$ denotes all indices of nodes $j$ connecting to node $i$ with nonzero edge features. $Z$ denotes the size of latent edge features.

The graph embedding operator is defined as follow:
\begin{equation}
\mathbf{\mathcal{G}^Z} =  \sum_{i=1}^{N}  \mathbf{\Theta} \mathbf{x}_{i}
\label{Eq:graphembed}
\end{equation}
Where $\mathcal{G}^Z$ denotes the graph embedding of size $Z$ all nodes of the graph.  $ \mathbf{\Theta}$ denote the trainable network weights. $N$ denotes the number of nodes in graph. $Z$ denotes the size of latent node features.

 Random edge drop was applied to augment data during training. The weighted loss was applied in the network to mitigate the effect of data imbalance. Learning rate decay was used to stabilize the training process. 
 Early stopping mechanism, weight decay, and dropout layers after fully connected layers were used to prevent over-fitting.

\subsection{Benchmark models}
\begin{figure}[h]
\begin{center}
\centerline{\includegraphics[width=\columnwidth]{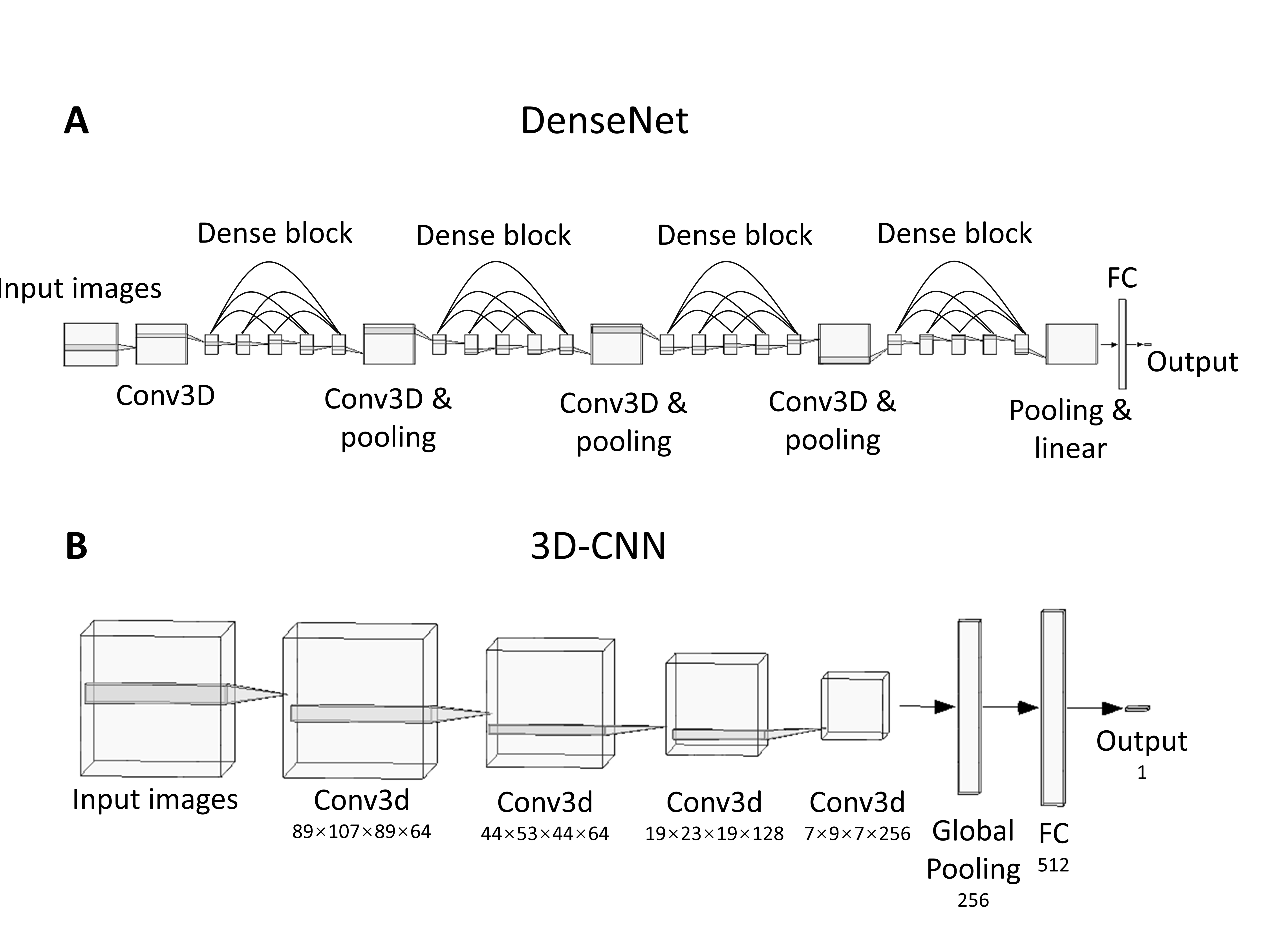}}
\caption{Architecture of the benchmark models. \textbf{A}. Classic three-dimensional Densely Connected Convolutional Networks (3D-DenseNet) consist of four convolutional layers and four densely connected blocks. \textbf{B}. Three-dimensional convolutional neural networks (3D-CNN) consist of four hidden convolutional layers with max-pooling and batch normalization, one global pooling layer followed by dropout, and one fully connected dense layer.}
\label{Fig:benchmark}
\end{center}
\end{figure}

We adopted a three-dimensional Densely Connected Convolutional Networks (3D-DenseNet)(Figure \ref{Fig:benchmark}A) and a three-dimensional convolutional neural networks (3D-CNN) (Figure \ref{Fig:benchmark}B)  as the benchmarks. Specifically, a classic 121-layer version of 3D-DenseNet follows the architecture described in~\cite{liang2018multimodal} while a traditional 3D-CNN with four hidden convolutional layers with batch normalization and pooling was applied, followed by a max-pooling layer and an output layer. Data were split using the same method as the GNN model. Weighted loss, learning rate decay, and early stopping are all applied, which was similar to the GNN  settings. The same loss function and optimiser were applied to the benchmark models as the GNN  model. Two experiments with different input were conducted: whole-brain MRI and MRI voxels inside tumor ROIs (contrasting-enhancing tumor core and necrosis) which are generated according to \cite{bakas2017advancing}.

\section{Results and Discussion}
\subsection{Model performance}

Our experiments show that the proposed model performs better than the baseline models (Table 1) for both cross-validation and testing. Interestingly, the benchmark models with tumor voxels as inputs perform better than the models with the whole brain as inputs, which suggests the potential bias from the extensive brain regions beyond the local tumor. Of note, our proposed GNN model, leveraging the brain network generated based on prior atlas and whole brain MRI, performs better than all the benchmark models, which may suggest that incorporating prior knowledge of brain networks could help the deep learning models capture more informative features regarding tumor invasion over either the local tumor or the whole brain.

\begin{table}[H]
\caption{Performances of IDH prediction models}
\label{tab:my-table}
\resizebox{\textwidth}{!}{%
\begin{tabular}{cccc}
\hline
\textbf{Methods} & \textbf{Accuracy $\left(\%\right)$}  & \textbf{Sensitivity $\left(\%\right)$} & \textbf{Specificity $\left(\%\right)$}  \\ \hline
\multicolumn{4}{c}{Cross-validation} \\ \hline
3D-CNN + whole brain MRI & 69.1 & 61.2 & 72.1 \\
3D-CNN + tumor ROIs & 80.1 & 77.7 & 81.0 \\
3D-DenseNet + whole brain MRI & 76.1 & 67.0 & 79.6 \\
3D-DenseNet + tumor ROIs & 84.1 & 86.4 & 83.3 \\
GNN + brain networks & 87.9 & 97.4 & 88.1 \\ \hline
\multicolumn{4}{c}{Test} \\ \hline
3D-CNN + whole brain MRI & 67.2 & 63.1 & 68.8 \\
3D-CNN + tumor ROIs & 78.2 & 75.7 & 79.2 \\
3D-DenseNet + whole brain MRI & 73.1 & 63.1 & 77.0 \\
3D-DenseNet + tumor ROIs & 83.3 & 83.5 & 83.2 \\
GNN + brain networks & 86.6 & 87.7 & 86.3 \\ \hline
\end{tabular}%
}
\end{table}

\subsection{Model interpretation}
To interpret the learning process of the GNN model, we applied the GNNExplainer~\cite{ying2019gnnexplainer}. GNNExplainer outputs a probability score that infers the importance of the edges in the prediction task and outputs a compact subnetwork of the networks. The task was achieved by maximizing both a graph neural network’s prediction and distribution of possible subnetworks. Only subnetworks with edges that have probability scores greater than $50\%$ were retained.

\begin{figure}[H]
\begin{center}
\centerline{\includegraphics[width=\columnwidth]{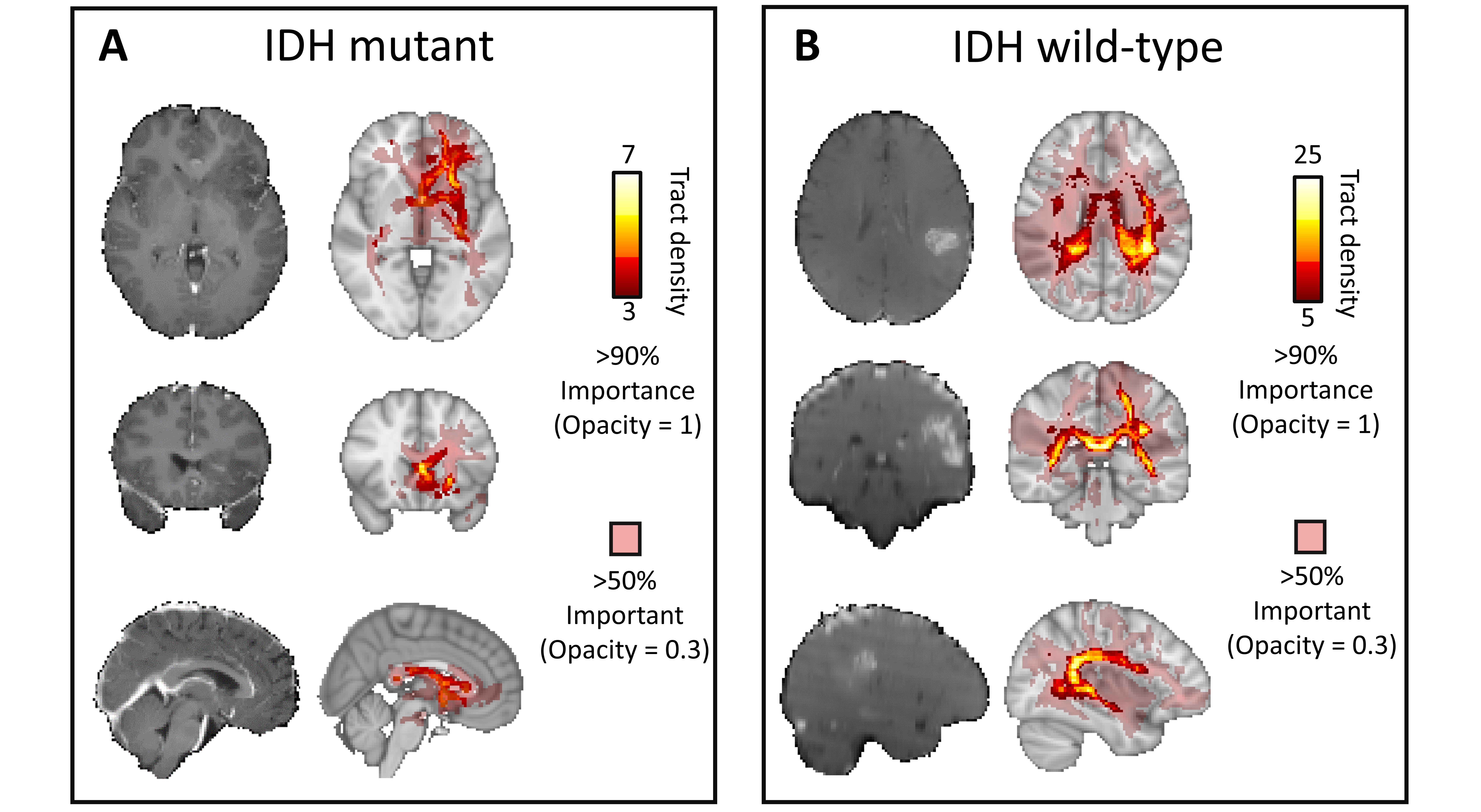}}
\caption{Examples of IDH mutant and wild-type.  \textbf{A} IDH mutant  \textbf{B}. IDH wild-type. For both patients, the left panels indicate the T1-weighted images and the right panels show the output of GNNExplainer, illustrating the voxel distribution of edges that have over 50\%  and 90\% probability of importance in IDH mutation prediction. The tract density of a voxel is defined as the number of tracts crossing the voxel. }
\label{Fig:GNNExplainer}
\end{center}
\end{figure}

Overall, we observe that the IDH wild-type is associated with a wider distribution of edge invasion, captured by the GNN model. Figure ~\ref{Fig:GNNExplainer} presents two typical cases of IDH mutant and wild-type, respectively, which also present the distribution of key white matter tracts (edges) that are important to the prediction accuracy.  In line with our prior knowledge that IDH wild-type generally causes more widespread invasion, the results of the model interpretation could further support the usefulness the proposed GNN model.

\section{Conclusion}
In this paper, we propose a method to generate brain networks based on multi-modal MRI and predict the IDH mutation status using GNN and the generated brain networks. Numerical results demonstrate that the proposed method outperforms benchmark methods. In future work, we could use the radiomic approach to extract representative features from the node and edge ROIs. Furthermore, special end-to-end GNN models could be developed to directly take the high dimensional multi-modal MRI voxels as inputs. To conclude, combining brain networks with GNN promises to serve as a novel powerful tool for deep learning model development in radiogenomic studies. 


%

\bibliography{reference.bib}

\begin{thebibliography}{10}
\providecommand{\url}[1]{\texttt{#1}}
\providecommand{\urlprefix}{URL }
\providecommand{\doi}[1]{https://doi.org/#1}

\bibitem{avants2009advanced}
Avants, B.B., Tustison, N., Song, G., et~al.: Advanced normalization tools
  (ants). Insight j  \textbf{2}(365),  1--35 (2009)

\bibitem{bakas2017advancing}
Bakas, S., Akbari, H., Sotiras, A., Bilello, M., Rozycki, M., Kirby, J.S.,
  Freymann, J.B., Farahani, K., Davatzikos, C.: Advancing the cancer genome
  atlas glioma mri collections with expert segmentation labels and radiomic
  features. Scientific data  \textbf{4}(1),  1--13 (2017)

\bibitem{bullmore2011brain}
Bullmore, E.T., Bassett, D.S.: Brain graphs: graphical models of the human
  brain connectome. Annual review of clinical psychology  \textbf{7},  113--140
  (2011)

\bibitem{choi2021fully}
Choi, Y.S., Bae, S., Chang, J.H., Kang, S.G., Kim, S.H., Kim, J., Rim, T.H.,
  Choi, S.H., Jain, R., Lee, S.K.: Fully automated hybrid approach to predict
  the idh mutation status of gliomas via deep learning and radiomics.
  Neuro-oncology  \textbf{23}(2),  304--313 (2021)

\bibitem{fagerholm2015disconnection}
Fagerholm, E.D., Hellyer, P.J., Scott, G., Leech, R., Sharp, D.J.:
  Disconnection of network hubs and cognitive impairment after traumatic brain
  injury. Brain  \textbf{138}(6),  1696--1709 (2015)

\bibitem{hyare2019modelling}
Hyare, H., Rice, L., Thust, S., Nachev, P., Jha, A., Milic, M., Brandner, S.,
  Rees, J.: Modelling mr and clinical features in grade ii/iii astrocytomas to
  predict idh mutation status. European journal of radiology  \textbf{114},
  120--127 (2019)

\bibitem{jenkinson2012fsl}
Jenkinson, M., Beckmann, C.F., Behrens, T.E., Woolrich, M.W., Smith, S.M.: Fsl.
  Neuroimage  \textbf{62}(2),  782--790 (2012)

\bibitem{jenkinson2001global}
Jenkinson, M., Smith, S.: A global optimisation method for robust affine
  registration of brain images. Medical image analysis  \textbf{5}(2),
  143--156 (2001)

\bibitem{li2019multi}
Li, C., Wang, S., Serra, A., Torheim, T., Yan, J.L., Boonzaier, N.R., Huang,
  Y., Matys, T., McLean, M.A., Markowetz, F., et~al.: Multi-parametric and
  multi-regional histogram analysis of mri: modality integration reveals
  imaging phenotypes of glioblastoma. European radiology  \textbf{29}(9),
  4718--4729 (2019)

\bibitem{li2019characterizing}
Li, C., Wang, S., Yan, J.L., Torheim, T., Boonzaier, N.R., Sinha, R., Matys,
  T., Markowetz, F., Price, S.J.: Characterizing tumor invasiveness of
  glioblastoma using multiparametric magnetic resonance imaging. Journal of
  neurosurgery  \textbf{132}(5),  1465--1472 (2019)

\bibitem{liang2018multimodal}
Liang, S., Zhang, R., Liang, D., Song, T., Ai, T., Xia, C., Xia, L., Wang, Y.:
  Multimodal 3d densenet for idh genotype prediction in gliomas. Genes
  \textbf{9}(8), ~382 (2018)

\bibitem{liu2020altered}
Liu, Y., Yang, K., Hu, X., Xiao, C., Rao, J., Li, Z., Liu, D., Zou, Y., Chen,
  J., Liu, H.: Altered rich-club organization and regional topology are
  associated with cognitive decline in patients with frontal and temporal
  gliomas. Frontiers in human neuroscience  \textbf{14}, ~23 (2020)

\bibitem{louis20162016}
Louis, D.N., Perry, A., Reifenberger, G., Von~Deimling, A., Figarella-Branger,
  D., Cavenee, W.K., Ohgaki, H., Wiestler, O.D., Kleihues, P., Ellison, D.W.:
  The 2016 world health organization classification of tumors of the central
  nervous system: a summary. Acta neuropathologica  \textbf{131}(6),  803--820
  (2016)

\bibitem{morris2019weisfeiler}
Morris, C., Ritzert, M., Fey, M., Hamilton, W.L., Lenssen, J.E., Rattan, G.,
  Grohe, M.: Weisfeiler and leman go neural: Higher-order graph neural
  networks. In: Proceedings of the AAAI Conference on Artificial Intelligence.
  vol.~33, pp. 4602--4609 (2019)

\bibitem{nyul2000new}
Ny{\'u}l, L.G., Udupa, J.K., Zhang, X.: New variants of a method of mri scale
  standardization. IEEE transactions on medical imaging  \textbf{19}(2),
  143--150 (2000)

\bibitem{ostrom2016cbtrus}
Ostrom, Q.T., Gittleman, H., Xu, J., Kromer, C., Wolinsky, Y., Kruchko, C.,
  Barnholtz-Sloan, J.S.: Cbtrus statistical report: primary brain and other
  central nervous system tumors diagnosed in the united states in 2009--2013.
  Neuro-oncology  \textbf{18}(suppl\_5),  v1--v75 (2016)

\bibitem{tcgalgg}
{Pedano, N., Flanders, A. E., Scarpace, L., Mikkelsen, T., Eschbacher, J. M.,
  Hermes, B., … Ostrom, Q}: {Radiology Data from The Cancer Genome Atlas Low
  Grade Glioma [TCGA-LGG] collection. The Cancer Imaging Archive.} (2016),
  \url{http://doi.org/10.7937/K9/TCIA.2016.L4LTD3TK}

\bibitem{rice2017lessp}
Price, S.J., Allinson, K., Liu, H., Boonzaier, N.R., Yan, J.L., Lupson, V.C.,
  Larkin, T.J.: Less invasive phenotype found in isocitrate
  dehydrogenase--mutated glioblastomas than in isocitrate dehydrogenase
  wild-type glioblastomas: a diffusion-tensor imaging study. Radiology
  \textbf{283}(1),  215--221 (2017)

\bibitem{salvalaggio2020post}
Salvalaggio, A., De~Filippo De~Grazia, M., Zorzi, M., Thiebaut~de Schotten, M.,
  Corbetta, M.: Post-stroke deficit prediction from lesion and indirect
  structural and functional disconnection. Brain  \textbf{143}(7),  2173--2188
  (2020)

\bibitem{tcgagbm}
{Scarpace, L., Mikkelsen, T., Cha, S., Rao, S., Tekchandani, S., Gutman, D.,
  Saltz, J. H., Erickson, B. J., Pedano, N., Flanders, A. E., Barnholtz-Sloan,
  J., Ostrom, Q., Barboriak, D., and Pierce, L. J.}: {Radiology Data from The
  Cancer Genome Atlas Glioblastoma Multiforme [TCGA-GBM] collection [Data set].
  The Cancer Imaging Archive} (2016),
  \url{https://doi.org/10.7937/K9/TCIA.2016.RNYFUYE9}

\bibitem{ivygap}
{Shah, N., Feng, X., Lankerovich, M., Puchalski, R. B., and Keogh, B. }: {Data
  from Ivy GAP [Data set]. The Cancer Imaging Archive.} (2016),
  \url{https://doi.org/10.7937/K9/TCIA.2016.XLWAN6NL}

\bibitem{smith2002fast}
Smith, S.M.: Fast robust automated brain extraction. Human brain mapping
  \textbf{17}(3),  143--155 (2002)

\bibitem{smith1997susan}
Smith, S.M., Brady, J.M.: Susan—a new approach to low level image processing.
  International journal of computer vision  \textbf{23}(1),  45--78 (1997)

\bibitem{stoecklein2020resting}
Stoecklein, V.M., Stoecklein, S., Gali{\`e}, F., Ren, J., Schmutzer, M.,
  Unterrainer, M., Albert, N.L., Kreth, F.W., Thon, N., Liebig, T., et~al.:
  Resting-state fmri detects alterations in whole brain connectivity related to
  tumor biology in glioma patients. Neuro-oncology  \textbf{22}(9),  1388--1398
  (2020)

\bibitem{tzourio2002automated}
Tzourio-Mazoyer, N., Landeau, B., Papathanassiou, D., Crivello, F., Etard, O.,
  Delcroix, N., Mazoyer, B., Joliot, M.: Automated anatomical labeling of
  activations in spm using a macroscopic anatomical parcellation of the mni mri
  single-subject brain. Neuroimage  \textbf{15}(1),  273--289 (2002)

\bibitem{wang2019invasion}
Wang, J., Xu, S.L., Duan, J.J., Yi, L., Guo, Y.F., Shi, Y., Li, L., Yang, Z.Y.,
  Liao, X.M., Cai, J., et~al.: Invasion of white matter tracts by glioma stem
  cells is regulated by a notch1--sox2 positive-feedback loop. Nature
  neuroscience  \textbf{22}(1),  91--105 (2019)

\bibitem{wei2021structural}
Wei, Y., Li, C., Cui, Z., Mayrand, R.C., Zou, J., Wong, A.L., Sinha, R., Matys,
  T., Sch{\"o}nlieb, C.B., Price, S.J.: Structural connectome quantifies tumor
  invasion and predicts survival in glioblastoma patients. bioRxiv  (2021)

\bibitem{wei2021quantifying}
Wei, Y., Li, C., Price, S.: Quantifying structural connectivity in brain tumor
  patients. medRxiv  (2021)

\bibitem{yan2009idh1}
Yan, H., Parsons, D.W., Jin, G., McLendon, R., Rasheed, B.A., Yuan, W., Kos,
  I., Batinic-Haberle, I., Jones, S., Riggins, G.J., et~al.: Idh1 and idh2
  mutations in gliomas. New England journal of medicine  \textbf{360}(8),
  765--773 (2009)

\bibitem{ying2019gnnexplainer}
Ying, R., Bourgeois, D., You, J., Zitnik, M., Leskovec, J.: Gnnexplainer:
  Generating explanations for graph neural networks. Advances in neural
  information processing systems  \textbf{32}, ~9240 (2019)

\end{thebibliography}
\end{document}